\RequirePackage{fix-cm}

\documentclass[pdftex,onecolumn,epjc3]{svjour3}  
\smartqed  
\pdfoutput=1 

\usepackage[left=2.5cm, right=2.5cm, top=2cm, bottom=2cm]{geometry}
\usepackage{multirow}
\usepackage{multicol}
\usepackage{subcaption}
\usepackage{float}
\usepackage{placeins}
\usepackage{comment}
\usepackage{hyperref}
\usepackage{xcolor}
\usepackage{amsmath}
\usepackage{amssymb}
\usepackage{graphicx}

\usepackage[T1]{fontenc} 

\usepackage[nottoc]{tocbibind}

\journalname{Eur. Phys. J. C}
\begin{document}

\title{Tachyonic dark energy- Constraints from current observations
}


\author{Ramanpreet Singh \thanksref{e1,addr1, addr3}
        \and
       Athul C. N. \thanksref{e2,addr2}
       \and H.K. Jassal
       \thanksref{e3,addr1}
}

\thankstext{e1}{e-mail: ph22043@iisermohali.ac.in}
\thankstext{e2}{e-mail: athulcn@iitk.ac.in}
\thankstext{e3}{e-mail: hkjassal@iisermohali.ac.in}


\institute{Indian Institute of Science Education and Research Mohali,
SAS Nagar, Mohali 140306, Punjab, India \label{addr1}
           \and
          Indian Institute of Technology Kanpur,
Kalyanpur, Kanpur, 208016,  Uttar Pradesh, India  \label{addr2}
           \and
          G.K.S.M. Government  College Tanda Urmar,
Hoshiarpur, 144212, Punjab, India \label{addr3}
}

\date{Received: date / Accepted: date}

\maketitle

\begin{abstract}
Recent observations from the DESI survey have reignited the debate on the true nature of dark energy, challenging the standard model of cosmology. The results suggest a preference for dynamical dark energy rather than a constant. Several recent analyses of DESI data indicate that the universe's expansion may not be accelerating in the way suggested by supernova based cosmology. Motivated by these studies, we investigated a tachyon type scalar field $\phi$ as a model for dark energy, assuming an exponential potential for the field and performed parameter estimation using MCMC techniques. Such a model offers solutions that have $w\sim-1$ and are decelerating without requiring a phantom like equation of state (EoS). The present day value of EoS parameter is treated as a free parameter. However, for the reference model, we fix its present value to $-1$. The analysis is carried out using Pantheon+ and BAO measurements from DESI. The results show that both types of datasets consistently predict a turnaround in the EoS, regardless of whether $w_{\phi0}$ is treated as a free parameter or fixed to $-1$. The corresponding deceleration parameter also exhibits a future turnaround for both datasets when $w_{\phi0}$ is free. However, in the reference model with $w_{\phi0}=-1$, the deceleration parameter instead approaches $-1$ asymptotically. A model comparison  
shows that the Pantheon+ dataset favors free $w_{\phi0}$, while BAO observations prefer $w_{\phi0} = -1$. This indicates a disagreement in the future evolution predicted by the two datasets within the tachyon dark energy model.

\end{abstract}

\section{Introduction}
The cosmological constant dark energy model \cite{peebles2003cosmological,amendola2010dark,bamba2012dark}, while remarkably successful in explaining several cosmological observations, is plagued with the fine tuning problem and several cosmological tensions, a comprehensive discussion of which can be found in \cite{di2025cosmoverse}.
This has led to investigations of a wide range of alternative models. 
These include modifications to the dark energy equation of state parameter as well as more radical approaches involving modifications to general relativity itself, suggesting that Einstein’s equations may not fully capture the true nature of gravity. 
Although the second class of approaches leads to modified gravity theories such as $f(R)$ gravity \cite{de2010f,sotiriou2010f,nojiri2017modified,nojiri2011unified}, MOND \cite{sanders1998cosmology,bekenstein2004relativistic}, and other methods \cite{das2024aspects,das2014suppressing}, the first class seeks to address cosmological tensions within the framework of Einstein’s general relativity. One such approach involves modifying the dark energy equation of state through various parametrizations, such as the Chevallier–Polarski–Linder (CPL) and logarithmic parametrizations \cite{chevallier2001accelerating,linder2003probing,linder2003exploring,jassal2005wmap,jassal2005observational,efstathiou1999constraining,feng2011new,aldering2002overview}, under the assumption that dark energy behaves as a perfect fluid in a Friedmann Robertson Walker (FRW) universe. 
The recent data release from DESI (Dark Energy Spectroscopic Instrument) shows that the Baryon Acoustic Oscillation (BAO) observations favour the CPL parameterisation of the dark energy equation of state over the standard cosmological constant model. 
When combined with CMB data, the results reveal an approximately $2.2\sigma$ deviation from the predictions of the $\Lambda$CDM model \cite{adame2025desi}, indicating that the $\Lambda$CDM framework may no longer be sufficient to fully explain the observations. A similar study finds a preference for dynamical dark energy using the CMB data combined with BAO measurements from DESI DR2 and supernova datasets from DESY5, Pantheon+, and Union3 \cite{li2025robust}.

It has been suggested that scalar fields can effectively model the required dark energy behaviour, naturally giving rise to a time-varying equation of state parameter. Notable examples include quintessence models (with $w > -1$) \cite{barreiro2000quintessence,tsujikawa2005reconstruction,jassal2009comparison} and phantom models (with $w < -1$) \cite{bouali2019cosmological,novosyadlyj2012cosmological,novosyadlyj2013quintessence}. 
In these models, the equation of state parameter is determined by the form of the scalar field potential. The potential $V(\phi)$ is not unique, and different choices can result in distinct cosmological evolutions. However, in quintessence-type dark energy models, when the field evolves in the slow-roll regime ($\dot{\phi} \ll V(\phi)$), the equation of state parameter approaches $-1$, causing the quintessence field to mimic the behaviour of a cosmological constant. 
Current observations do not exclude the possibility that the dark energy equation of state could be less than  $-1$ \cite{ade2016planck}. 
In such scenarios, phantom-type dark energy models may become relevant {\cite{phanthom_desi,mishra2025braneworld,hossain2025current}}. However, these models can lead to the emergence of a Big Rip singularity within a finite proper time.
Nevertheless, these scalar field models prove to be inadequate in simultaneously addressing the cosmic coincidence problem and the fine-tuning of initial conditions \cite{ji2024scalar}.

An alternative scalar field model, which is inspired by string theory, is the tachyonic dark energy model. 
The tachyon scalar field models have a dark energy equation of the state parameter lying within the range $-1 \leq w \leq 0$, with $w$ approaching zero in the distant past. 
In the context of tachyonic scalar fields, two particularly well studied potentials are the inverse square potential and the exponential potential, which are 'runaway' potentials. 
Constraints on the tachyon scalar field parameters for both of these potentials were studied using low-redshift data in \cite{singh2019low}. 
The exponential potential is particularly significant as it avoids the future event horizon that arises in the standard $\Lambda$CDM model \cite{bagla2003cosmology,copeland2005needed,calcagni2006tachyon,rajvanshi2019reconstruction}. {Dark Energy Spectroscopic Instrument (DESI) results indicate that for $w_0-w_a$ model of dark energy, the combination of DESI data with either CMB or Type Ia supernova observations favors $w_0 > -1$ and $w_a < 0$, implying a stronger cosmic acceleration in the past \cite{adame2025desi,DESIDr2bao}}. Therefore
this model gains significance as the latest DESI data favours a slowing down of the acceleration of the Universe \cite{wang2025questioning,van2025suggestions,camphuis2025spt,son2025strong} and hence from a theoretical point of view, this model is a viable explanation.
Motivated by this, we perform a detailed analysis of this model and compare the evolution of the universe predicted with the latest observations.
Our results show that the  parameters in this model are consistent within  $1\sigma$. 
Additionally,  for a varying equation of state parameter, there is no tension in the Hubble constant ($H_0$) and other cosmological parameters \cite{Di_Valentino_2021,DAHMANI2023101266}.
Our analysis reveals a turnaround in both the equation of state parameter $w$ and the deceleration parameter $q$ in the past;  the current accelerated expansion of the universe may eventually slow down or go back to the decelerated expansion in the future. 

This paper is structured as follows: Section \ref{Theory} presents the theoretical framework of the tachyon model with an exponential potential in a homogeneous and isotropic universe. Section \ref{Data} describes the observational data and the methodology used in our analysis. Finally, Section \ref{summary} provides the summary and conclusions of this analysis.

\section{Tachyon Cosmology}
\label{Theory}
A homogeneous and isotropic universe is described by the Friedmann-Lemaitre-Robertson-Walker (FLRW) metric, the maximally symmetric solution to Einstein's field equations. 
We assume a spatially flat universe with matter and radiation modelled as perfect fluids.
In a flat FLRW universe filled with matter, radiation and dark energy described by a scalar field, the Friedman equations are 
\begin{align}
   & \left( \frac{\dot{a}}{a}\right)^2 =\frac{8\pi G}{3}\left( \rho_{m}+\rho_{r} +\rho_{\phi}\right)
    \\
    & \frac{\ddot{a}}{a}
    =-\frac{4\pi G}{3} \left(\rho_\phi +3P_\phi+\rho_r +3P_r
    +\rho_{{m}}\right)
    \\
     & \frac{\ddot{\phi}}{1-\dot{\phi}^2}+3H\dot{\phi}+\frac{1}{V}\frac{dV(\phi)}{d\phi} =0
    \label{Friedman}
\end{align}
\\
The tachyon scalar field is described by the Lagrangian {\cite{singh2019low,bagla2003cosmology}} 
\begin{equation}
    \mathcal{L} = -V(\phi) \sqrt{1 - \partial_\mu \phi \partial^\mu \phi}
    \label{lagrangian}
\end{equation} 
where \( V(\phi) \) is an arbitrary potential. 
The energy density, pressure, and the equation of state parameter of the tachyon field are  
\begin{align}
    \rho_\phi &= \frac{V(\phi)}{\sqrt{1 - \dot{\phi}^2}}
    \\
    P_\phi &= -V(\phi) \sqrt{1 - \dot{\phi}^2}
    \\
    \Rightarrow  w_\phi &= \frac{P_\phi}{\rho_\phi} = \dot{\phi}^2 - 1
    \label{density}
\end{align}
This suggests that $w_\phi\geq -1$ since $\dot{\phi}^2\geq 0$.

Tachyon fields can be effectively modelled by an inverse square potential, in this case we consider only the exponential potential.
These 'runaway' potentials have been extensively studied and their parameters have been constrained using different observations {\cite{singh2019low,bagla2003cosmology}}.
In these models, for some combinations of parameter values, the matter density may not dominate completely at high redshift. 
Therefore tachyon models have also been considered to describe dark matter as well as dark energy \cite{padmanabhan2002accelerated,bagla2003cosmology}. 
Although the inverse square potential is an effective, viable model, for the purpose of this  paper, we consider the exponential potential.
{The exponential  potential is of  significance, particularly in scalar field models of dark energy \cite{copeland2006dynamics}. It describes an accelerating phase that may later transition into a dust-like phase \cite{bagla2003cosmology}.}
Therefore, such a model may provide a plausible explanation for the `slowing' down of the expansion of the Universe as indicated by the DESI observations \cite{karim2025desi}. 
This model does not have a future horizon.
A comprehensive discussion of this can be found in the  reference \cite{bagla2003cosmology}.

We assume the tachyon field potential to be given by {\cite{singh2019low,bagla2003cosmology}}
\begin{align}
    V(\phi) = V_0 \exp\left(-\frac{\phi}{\phi_a}\right)
\end{align} 
where \( V_0 \) and \( \phi_a \) are the scalar field parameters.  
To solve the cosmological equations numerically, we transform the above equations by introducing the following dimensionless variables \cite{bagla2003cosmology}:  
\begin{equation}
    y = \frac{a(t)}{a(t_{\text{in}})}, \quad \psi = \frac{\phi(t)}{\phi(t_{\text{in}})}, \quad x = H_{\text{in}} t
\end{equation}
where $t_{in}$ is some initial instant of
 time {and $H_{in}$ is the Hubble parameter at time $t_{in}$}. The dimensionless density parameter of the tachyonic scalar field is

 \begin{align}
     \Omega_{\phi} &=\frac{\rho_{\phi}}{\rho_{critical}}
     =\frac{V_0 e^{-\phi/\phi_a}}{\sqrt{-w_\phi}}\frac{8\pi G}{3H^2}
 \end{align}
{where $G$ is the gravitational constant and $H$ is the Hubble parameter.} 
 This implies $w_{\phi}\leq 0$. Together with (\ref{density}), $-1\leq w_\phi\leq 0$. 
The present day value of the tachyon density \cite{singh2019low}
 \begin{align}
  \Rightarrow  V_0  \frac{8\pi G}{3H_0^2}
    &= \Omega_{\phi0} e^{\phi_0/\phi_a}\sqrt{-w_{\phi0}}
 \end{align}
Defining $\lambda=\frac{\phi_0}{\phi_a}$,
the evolution equations reduce to the following set of coupled differential equations
\begin{align}
 & y^\prime 
  =
  y\left[{(1-\Omega_{m0}-\Omega_{r0}) \frac{e^{-\lambda(\psi-1)}}{\left({1-H_0^2\phi^2_0 \psi^{\prime 2}}\right)^{\frac{1}{2}}} +\frac{\Omega_{m0}}{y^3}+ \frac{\Omega_{r0}}{y^4}}\right]^\frac{1}{2}
   \nonumber
  \end{align}
with the condition that  $\mid H_0 \phi_0 \psi^\prime \mid <1$,
  \begin{align}
  &
  \psi^{\prime\prime} =- \left(1-H_0^2\phi^2_0 \psi^{\prime 2} \right)
  \left( \frac{3}{y}y^\prime \psi^\prime -\frac{\lambda}{H_0^2 \phi^2_0}  \right) 
  \label{differ}
\end{align}
where the superscript prime denotes the derivative with respect to the variable $x$.
We solve the system of differential equations (\ref{differ}) and compare the  theoretical predictions with the  observational data.  
We choose the initial time $t_{in}=t_0$, corresponding to the present epoch. 
We define a dimensionless field variable $\psi=\phi/\phi_0$. 
As the subsequent analysis considers only data with \( z < 2.5 \), one finds that for fixed parameter values, the results are essentially insensitive to $x_0$, remaining nearly identical in the vicinity of $z = 0$. Therefore we set $x = x_0 = 0.7$ at the present time, noting that this choice does not affect the final results.
The initial conditions at $x=x_0$ are
     $y(x_0) = 1$,
    $\psi(x_0) = 1$,
 \begin{displaymath}
          \psi^\prime({x_0}) =  \frac{\sqrt{w_{\phi 0} + 1}}{\phi_0 H_0} 
    \end{displaymath}


From eq. \ref{differ}, it is evident that the parameters $H_0$ and $\phi_0$ appear only in the combination $H_0 \phi_0$, indicating a degeneracy between them, and we can define a new parameter $\delta = H_0 \phi_0$.
Furthermore, we neglect the contribution of radiation by setting $\Omega_r = 0$. 

We first evolve the equations back in time to enable comparison with the data. 
Once the model is calibrated in this way, we can then evolve the solution forward to explore the potential implications and predictions of the model. 
Solving the system with the specified initial conditions provides us with the evolution of $y \,(= a)$, $\psi$, and $\psi^\prime$ as functions of $x$ (or equivalently, $t$). 
We then convert $\psi$, $\psi^\prime$, and $x$ into functions of redshift $z$. 
Once we obtain $ \psi(z)$ and $\psi^\prime(z)$, we construct smooth interpolating functions for both the scalar field and its derivative as functions of redshift. 
These interpolators enable the evaluation of $E(z')$ and other relevant quantities required for comparison with observational data. 
This allows us to examine the deceleration parameter and its evolution with redshift for this model. We make use of the second Friedmann equation along with the relations
$ V_0 =\frac{3H_0^2}{8\pi G} (1-\Omega_{m0})e^\lambda, \quad 
    \phi/\phi_a =\psi\lambda, \quad
    \dot{\phi} =\delta \, \psi^{\prime }$
to derive the deceleration parameter in analytical form as follows:
\begin{align}
    q &= -\frac{\ddot{a}}{aH^2} =\frac{1}{2E(z)^2}\left[\frac{ (1-\Omega_{m0})e^{-\lambda(\psi-1)}}{\sqrt{1-\delta^2\psi^{\prime 2}}} \left(3\delta^2\psi^{\prime 2} -2\right)
    +\Omega_{{m0}}(1+z)^3\right]
\end{align}
{where $E(z) = H(z)/H_0$.}

\section{Data  and Methodology}
\label{Data}
We used the Pantheon plus (SNe data, referred to as `PANTHEON+') \cite{scolnic2022pantheon+}, Baryon Acoustic Oscillation (BAO)  data from \cite{PhysRevD93023530,deCarvalho2018,mnrasstaa119,Alcaniz2016ryy,2020APh11902432C,2011MNRAS4163017B,PhysRevD103083533,101093mnrasstv154,duMasdesBourboux2020} (henceforth referred to as BAO1) and the recent DESI data release 2 \footnote{{The Pantheon+ and DESI BAO datasets along with the covariance matrices are available in the cited references and via the repository https://github.com/CobayaSampler }}\cite{karim2025desi}. 
To perform the parameter estimation, we do the Markov Chain Monte Carlo (MCMC) analysis with 100 walkers, 3000 steps, and a burn-in phase of 500 steps. 
The chains were sufficiently long to satisfy the Gelman–Rubin convergence criterion, with $R < 1.01$.
It is important to note that, although $H_0$ and $\phi_0$ are degenerate at the level of the differential equations, this degeneracy is lifted when computing observable quantities such as the angular diameter distance $D_A$ and the luminosity distance $d_L$.

\subsection{DESI \& Earlier BAO Datasets }

The BAO datasets require the computation of the following key cosmological distance measures:

\begin{itemize}
    \item Angular Diameter Distance:
 \begin{equation}
      D_A(z) = \frac{1}{1 + z}  \frac{c}{H_0} \int_0^z \frac{dz'}{E(z')}
 \end{equation}

\item Volume-averaged Distance:
\begin{equation}
D_V(z) = \left[ (1 + z)^2 D_A(z)^2 \cdot \frac{c z}{H_0 E(z)} \right]^{1/3}
\end{equation}

\item Hubble Distance:
\begin{equation}
      D_H(z) = \frac{c}{H_0 E(z)}
\end{equation}

\item Transverse Comoving Distance:
\begin{equation}
     D_M(z) = \frac{c}{H_0} \int_0^z \frac{dz'}{E(z')}
\end{equation}

\item BAO Angular Scale:
\begin{equation}
     \theta(z) = \frac{r_s}{D_V(z)} \cdot \frac{180}{\pi}
\end{equation}
 
  where $r_s = 147.38 \, \text{Mpc}$ is the sound horizon scale \cite{2020}. 
 There exists a degeneracy between the Hubble constant $H_0$ and the sound horizon at the drag epoch $r_s$. However, the posterior distribution of $r_s$, as inferred by the Planck data, is sufficiently narrow to justify treating it effectively as fixed to this value within the uncertainties of the dataset considered, thereby breaking the degeneracy with $H_0$.
  

\end{itemize}

\subsection{Supernova Typa Ia Pantheon+ data}
For the Pantheon+ Supernovae (SNe) dataset, the analysis relies on the following key expressions:

\begin{itemize}
    \item Luminosity Distance:
\begin{equation}
     d_L(z) = (1 + z)  \frac{c}{H_0} \int_0^z \frac{dz'}{E(z')}
\end{equation}

\item Distance Modulus:
\begin{equation}
    \mu_{\text{th}}(z) = 5 \log_{10} \left( \frac{d_L(z)}{\text{Mpc}} \right) + 25  
\end{equation}

\end{itemize}
We use the standard $\chi^2$ statistics  to fit the model with data
\begin{equation}
  \chi^2 = \Delta X^T C^{-1} \Delta X 
\end{equation}
where $(\Delta X)_i=X_i  (observed) - X_i (theory)$ is the residual of the observable, $C$ denotes the covariance matrix of the data. 
The total $\chi^2$ from the BAO1 datasets as the sum of contributions from each observable as follows:
$
\chi^2 = \chi^2_{D_V/r_s} + \chi^2_{D_M/r_s} + \chi^2_{D_H/r_s} + \chi^2_{r_s/D_V} + \chi^2_{\theta_{\text{BAO}}}
$. For DESI DR2 data, the observables are $X$ are ${D_M/r_s},{D_H/r_s},{D_V/r_s}$ and $C$ is the covariance matrix associated with the DESI DR2 data. While for the Pantheon+ data, 
$
  \chi^2 = (\mu_{\text{obs}} - \mu_{\text{th}})^T C^{-1} (\mu_{\text{obs}} - \mu_{\text{th}})
$
  where $C$ is the covariance matrix associated with the Pantheon+ dataset.
We then use it to define the log-likelihood function and posterior probability, as given by Bayes' theorem, as:
\begin{align}
    \ln \mathcal{L} &= -\frac{1}{2} \chi^2
    \\
     \ln P(\theta \,|\, \text{data}) &= \ln \left[ \text{prior}(\theta) \right] + \ln \mathcal{L}
\end{align}

\subsection{Constraints}
\begin{figure}[ht]
\centering
\includegraphics[scale=0.35]{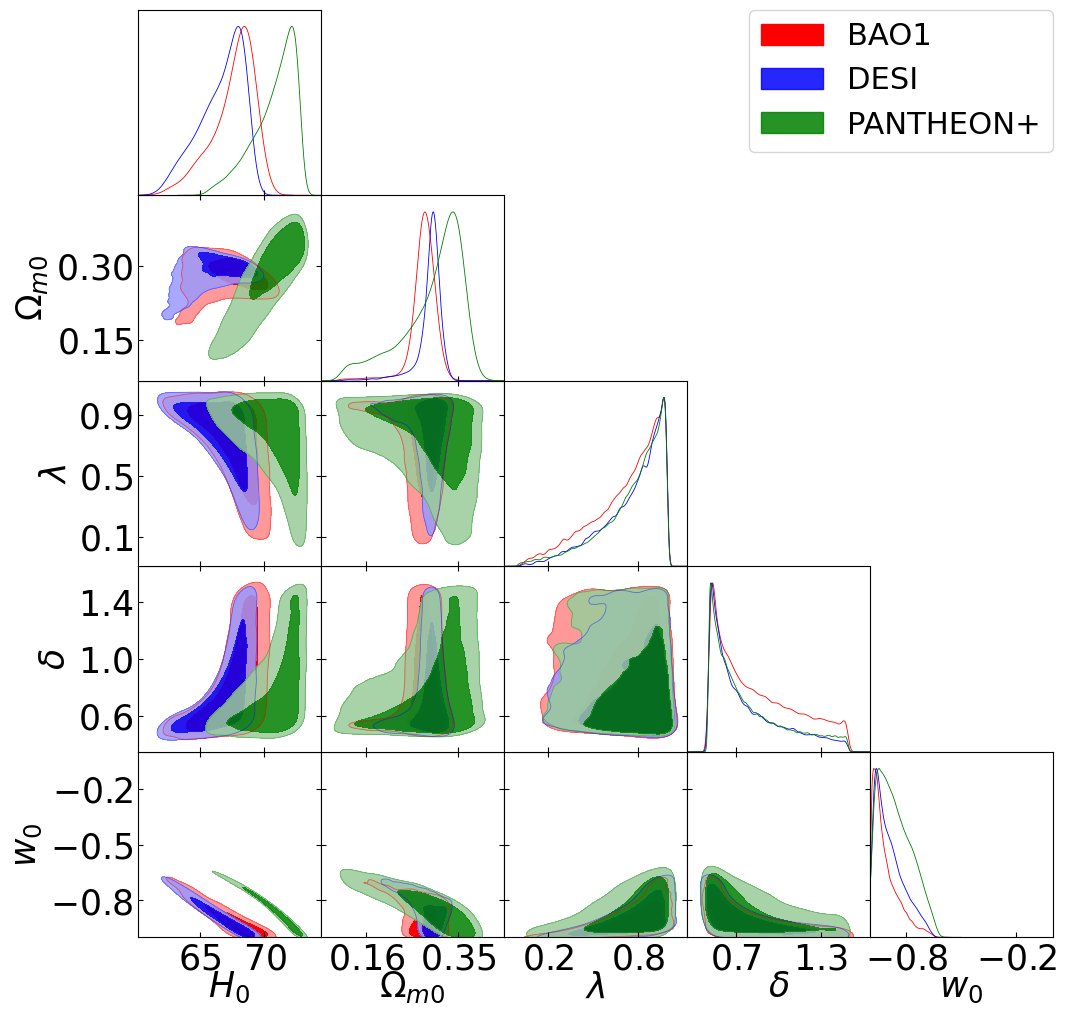} 
\caption{ Parameters at $2 \sigma$ confidence level for tachyonic scalar field model  with exponential potential. Here $w_{\phi 0}$ is a free parameter, and the constraints are obtained  via MCMC analysis with three different datasets i.e BAO (red), DESI (blue) and Pantheon+ (green). The uniform prior for MCMC is given in the table \ref{Param_bound w}. From the posterior plots, it is clear that the values of the parameters agree within the $\sim 1 \sigma$ range. {The solid coloured region corresponds to 1$\sigma$, while the lightly shaded region corresponds to 2$\sigma$ confidence level. } 
}
\label{mcmc w}
\end{figure}

\begin{figure}[ht]
\centering
\includegraphics[scale=0.35]{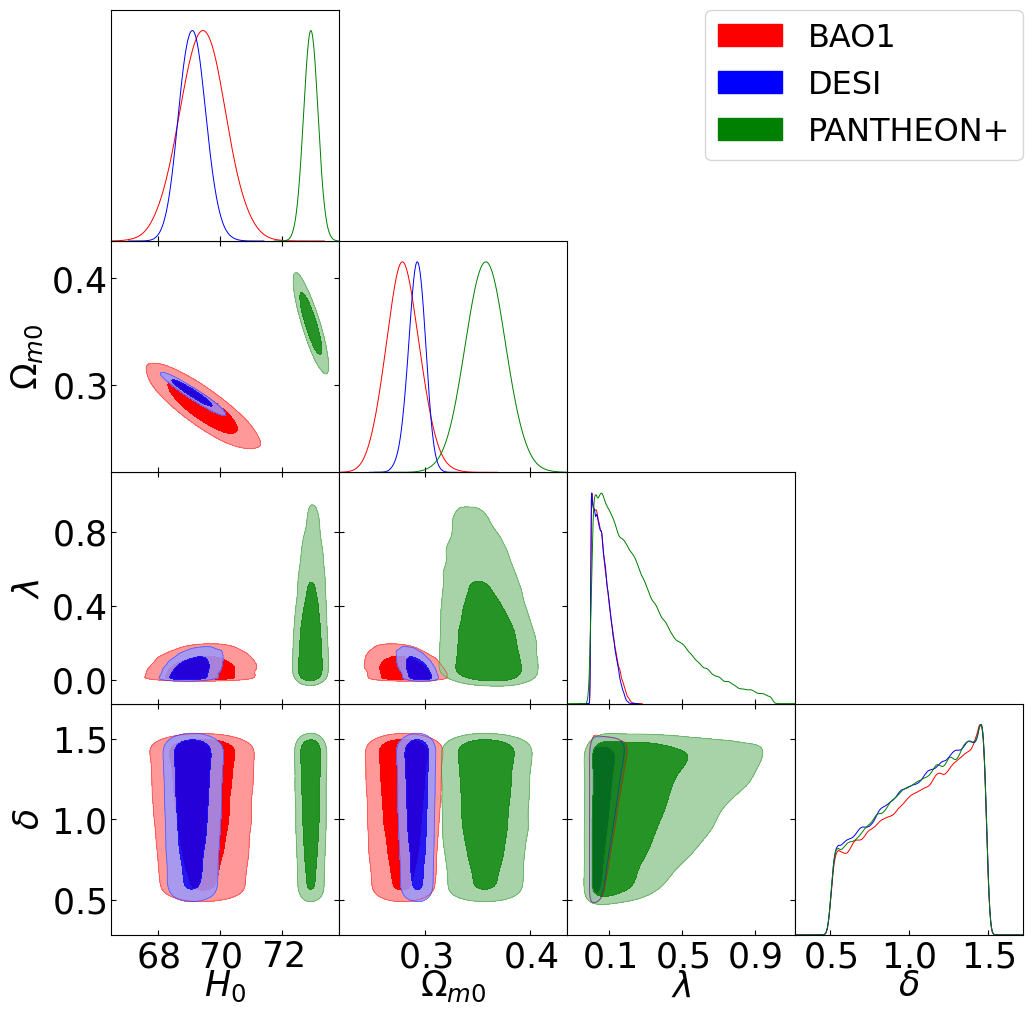} 
\caption{ Contours showing 2-$\sigma$ allowed region with $w_{\phi 0}=-1$. 
The uniform prior for MCMC is given in the table \ref{Param_bound no w}. {The colour scheme is same as fig \ref{mcmc w}. } 
}
\label{mcmc}
\end{figure}

\begin{figure}[h!]
\centering
\includegraphics[scale=0.22]{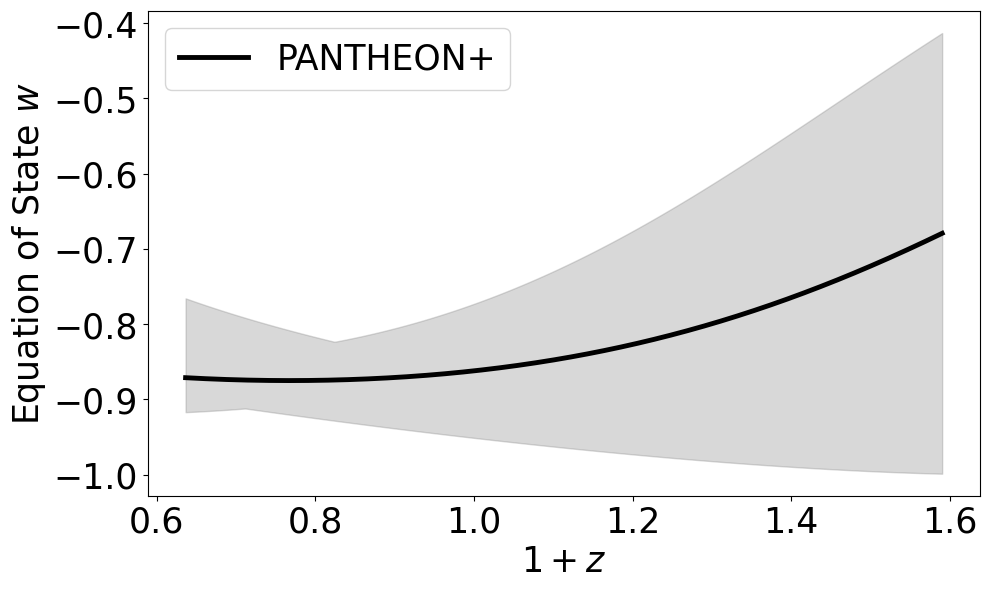}
\includegraphics[scale=0.22]{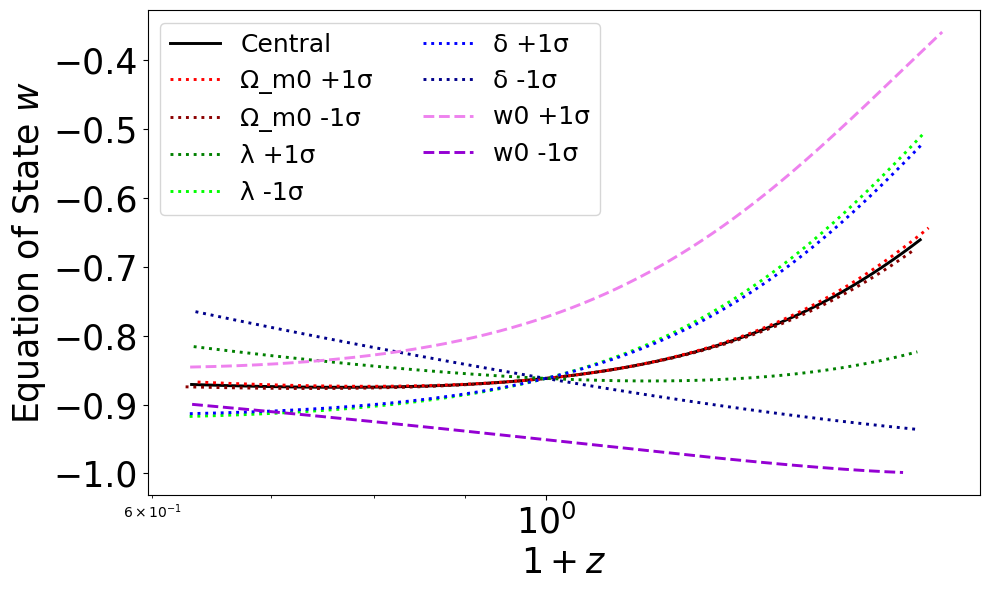}
\includegraphics[scale=0.22]{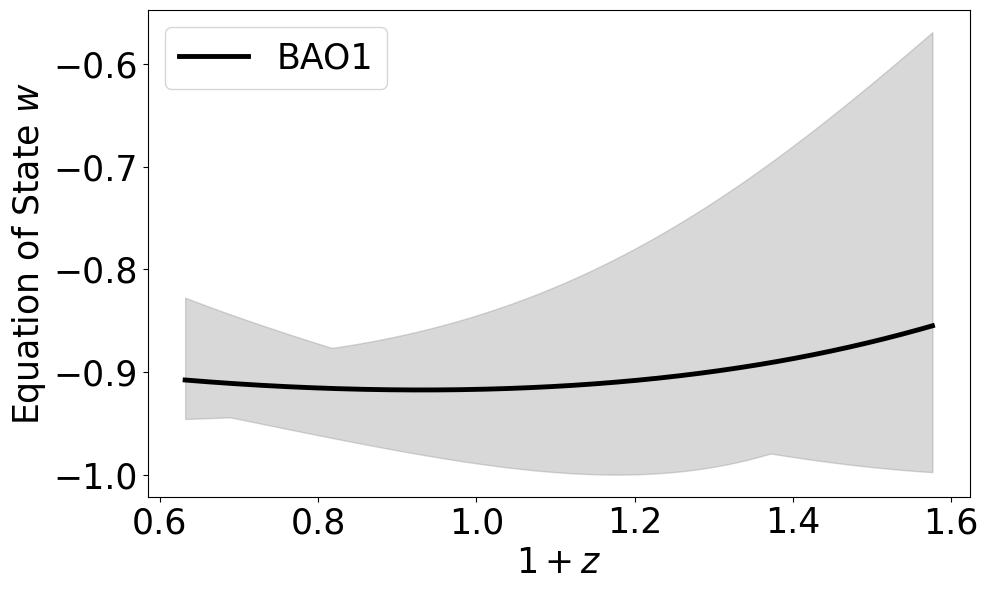}
\includegraphics[scale=0.22]{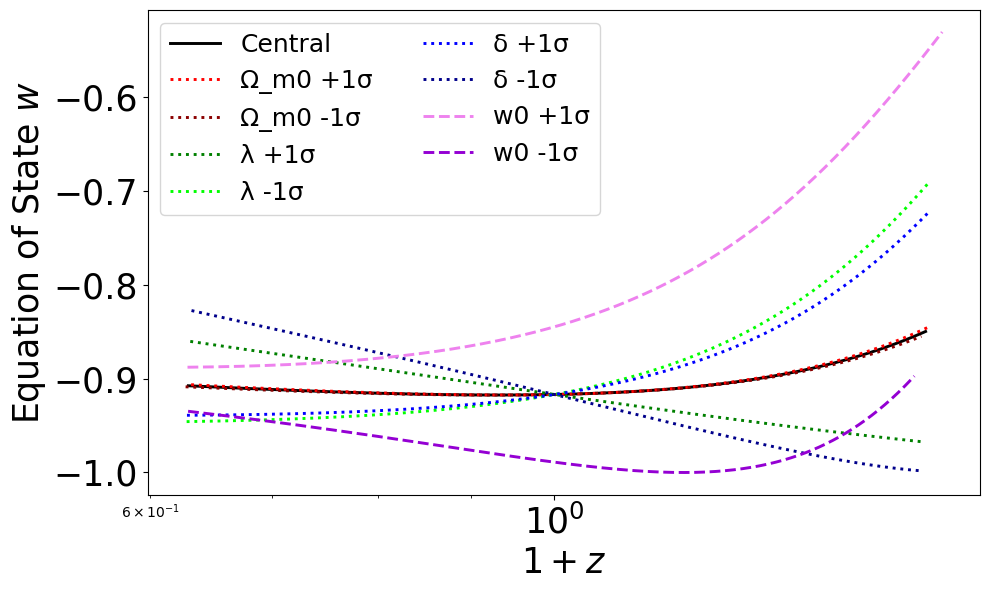}
\includegraphics[scale=0.22]{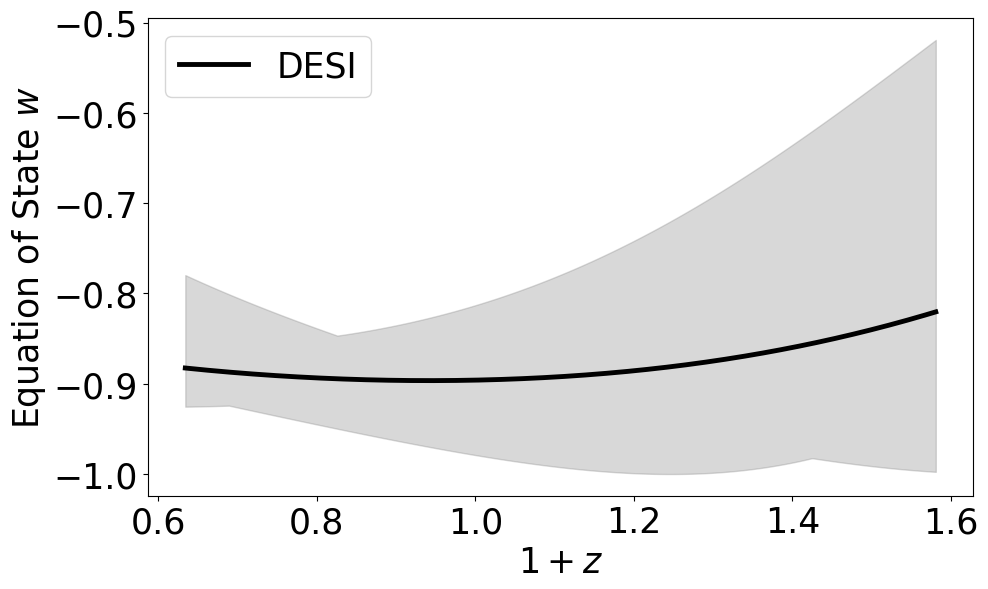}
\includegraphics[scale=0.22]{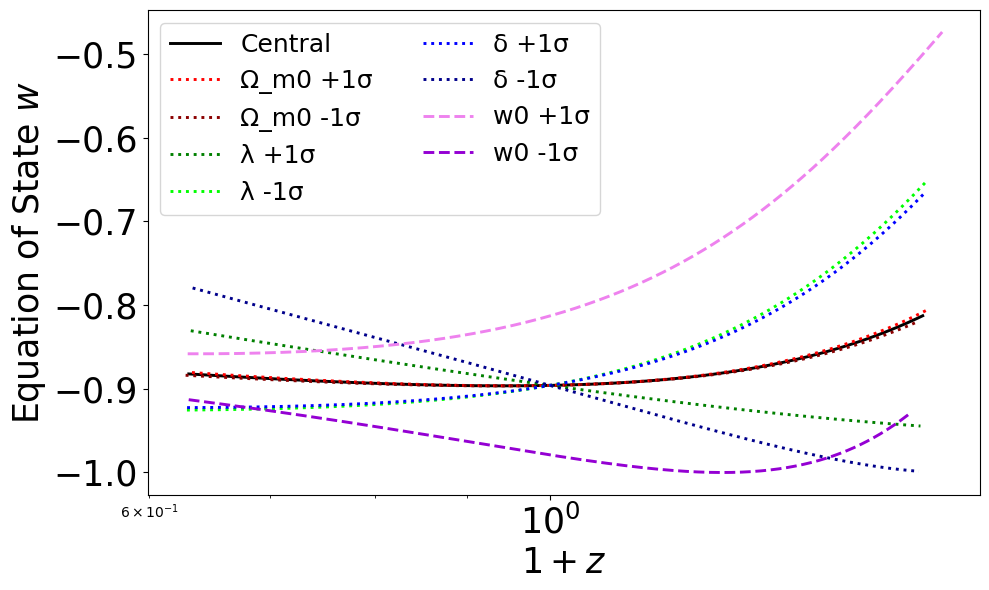}
\caption{This figure shows the equation of state parameter as a function of redshift. The dark curve represents the evolution of $w$ with redshift $z$ at the mean values of the parameters obtained from table \ref{Param_bound w}. The shaded region, referred to as the $1 \sigma$ region around this curve, represents the envelope formed by the outlier curves. The dotted curves correspond to scenarios where one parameter is varied by $1\sigma$, while the others are held fixed at their central values.
}
\label{eq of state w}
\end{figure}

\begin{figure}[h!]
\centering
\includegraphics[scale=0.22]{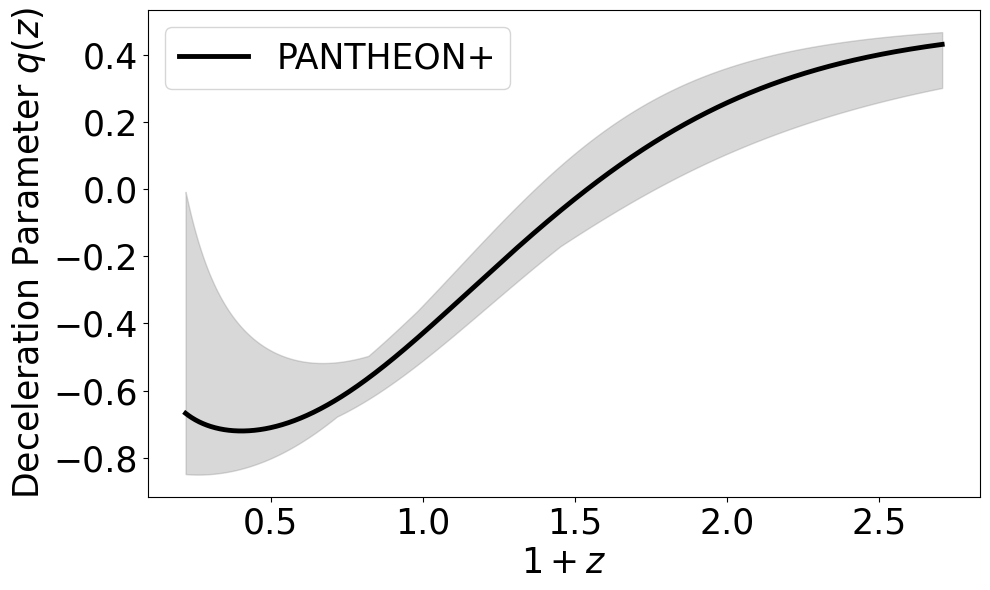}
\includegraphics[scale=0.22]{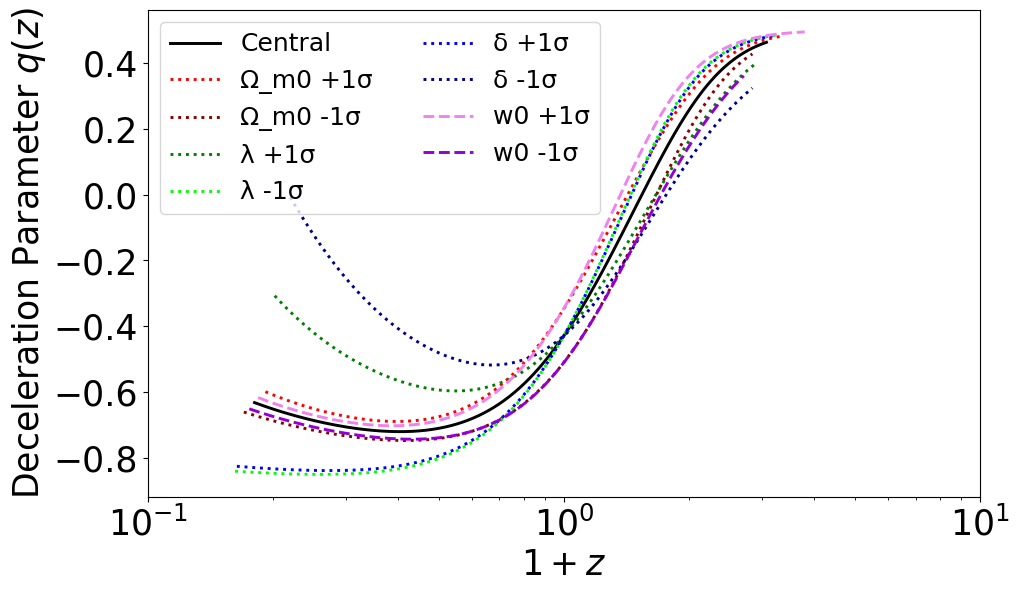}
\includegraphics[scale=0.22]{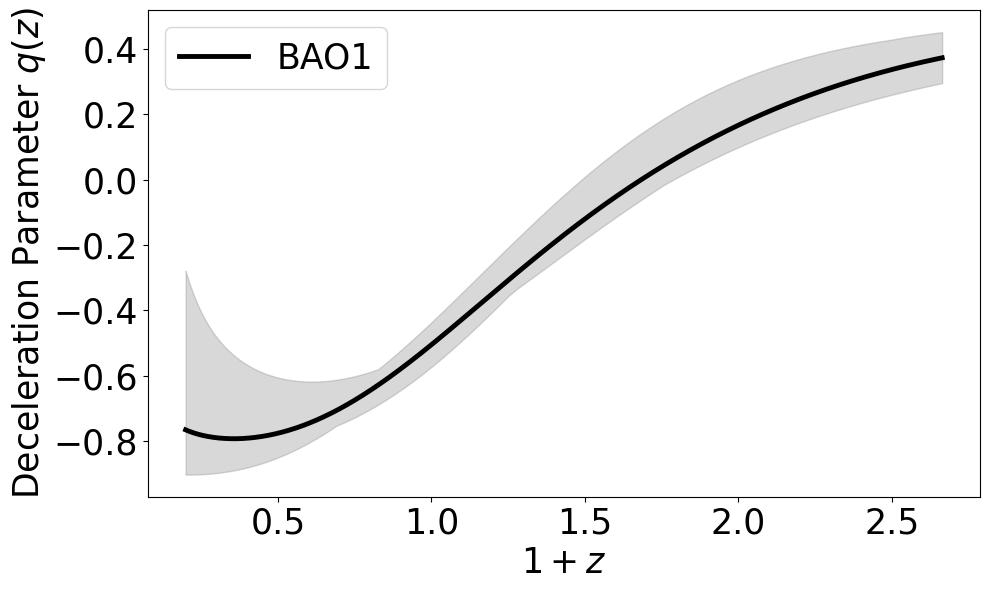}
\includegraphics[scale=0.22]{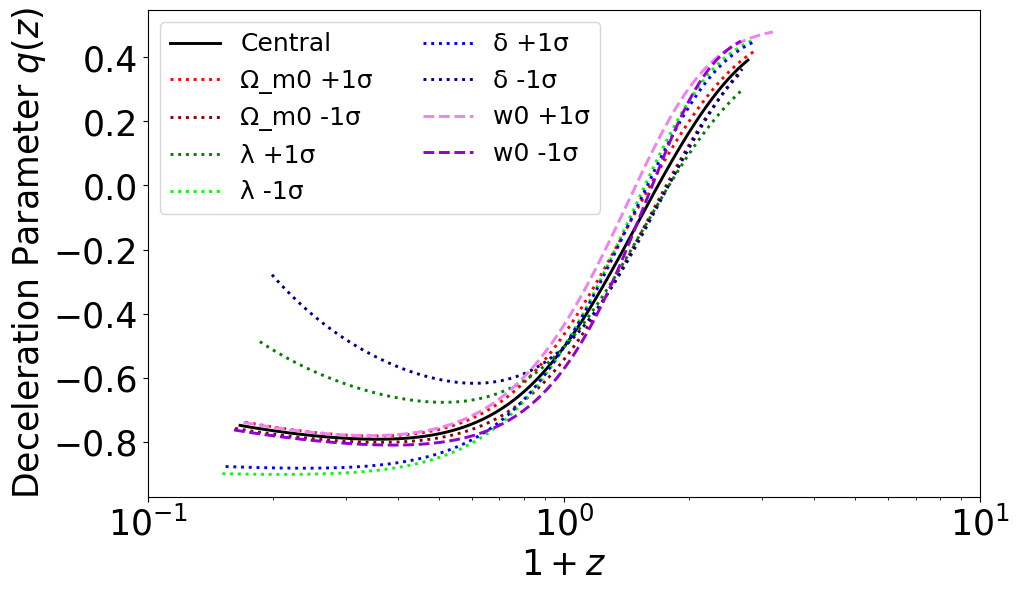}
\includegraphics[scale=0.22]{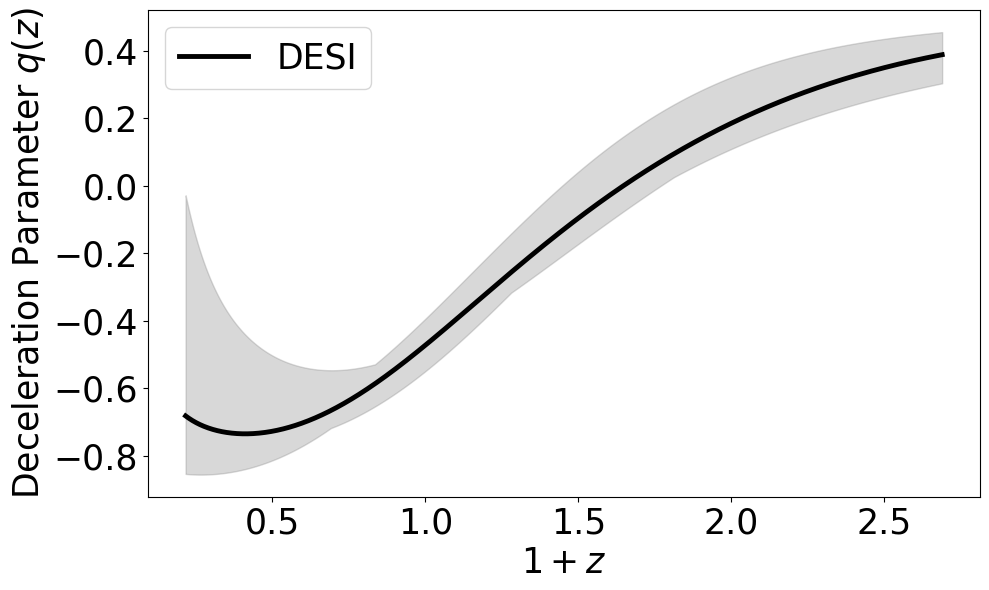}
\includegraphics[scale=0.22]{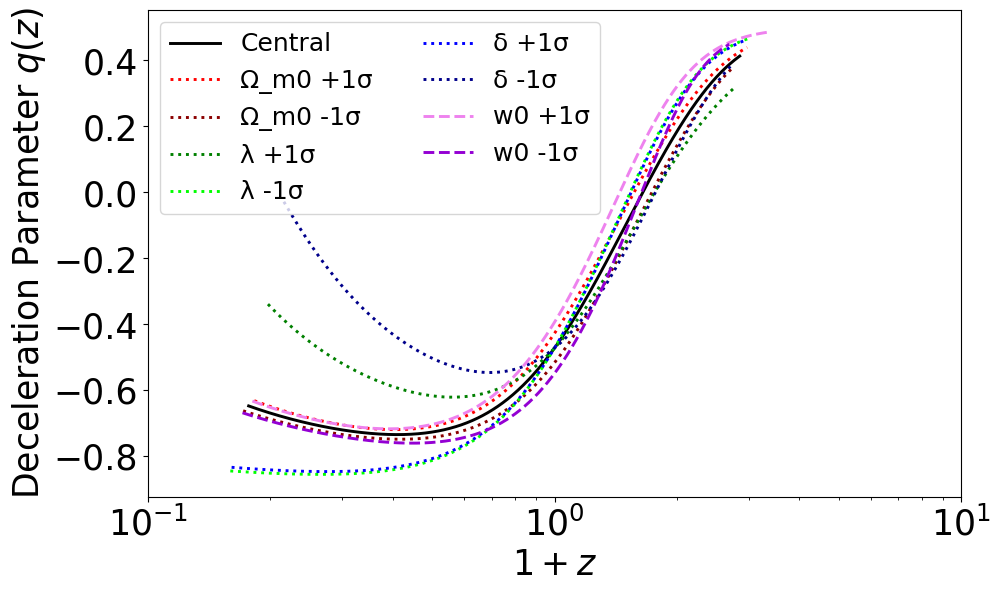}
\caption{Plots of deceleration parameter as a function of redshift. The dark curve represents the evolution of $q$ w.r.t. redshift $z$ at the mean values of the parameters obtained from table \ref{Param_bound w}.  This suggests that for all the datasets used, this tachyon model predicts the slowing down of acceleration in the future. }
\label{decelaration w}
\end{figure}

\begin{figure}[ht]
\centering
\includegraphics[scale=0.22]{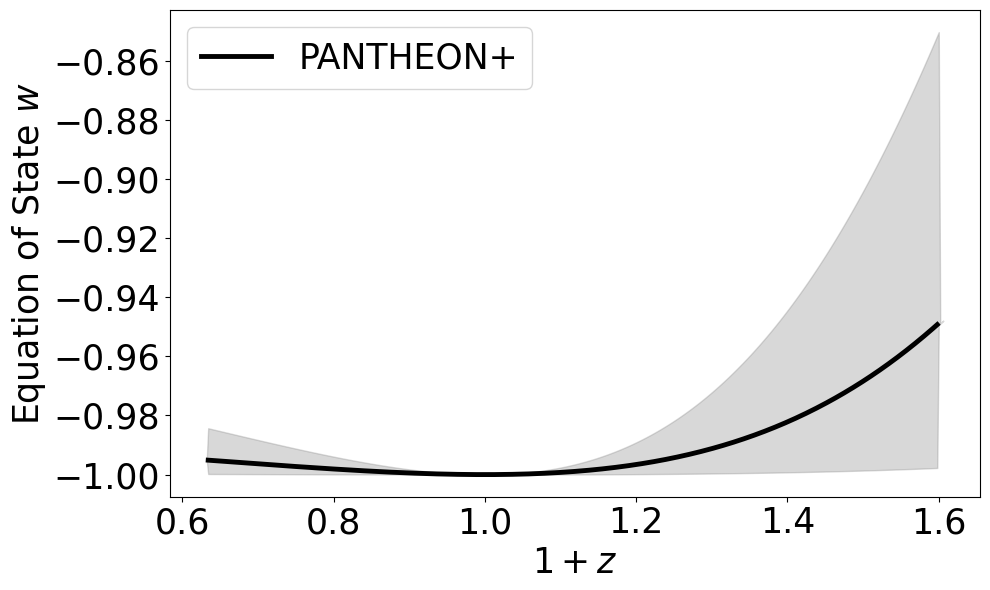}
\includegraphics[scale=0.22]{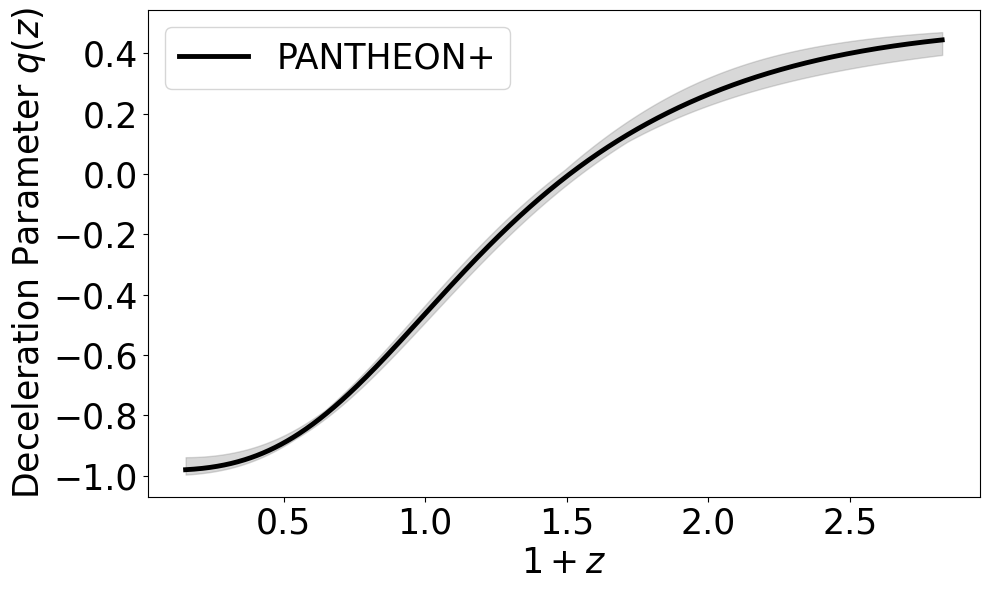}
\includegraphics[scale=0.22]{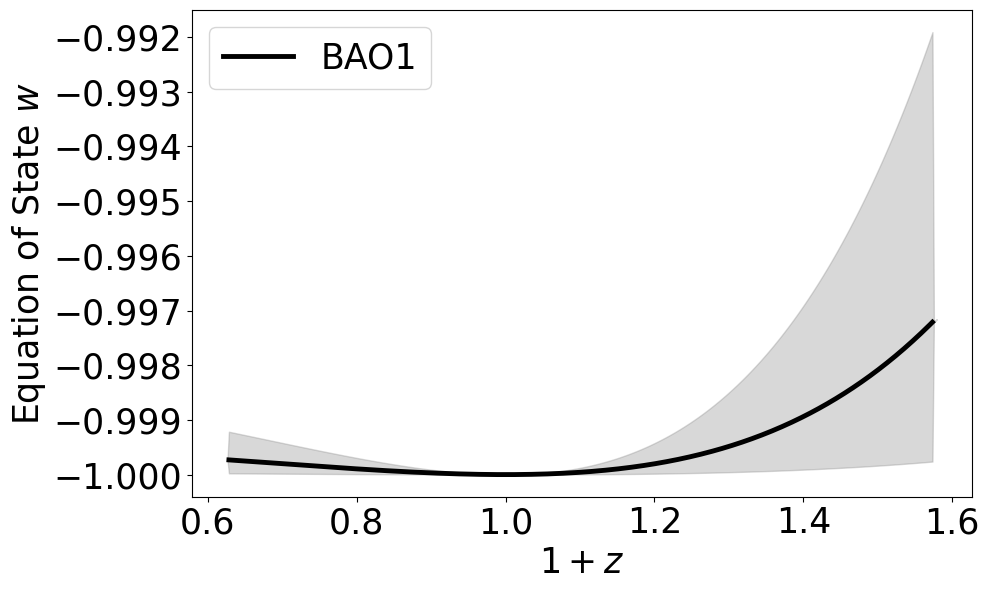}
\includegraphics[scale=0.22]{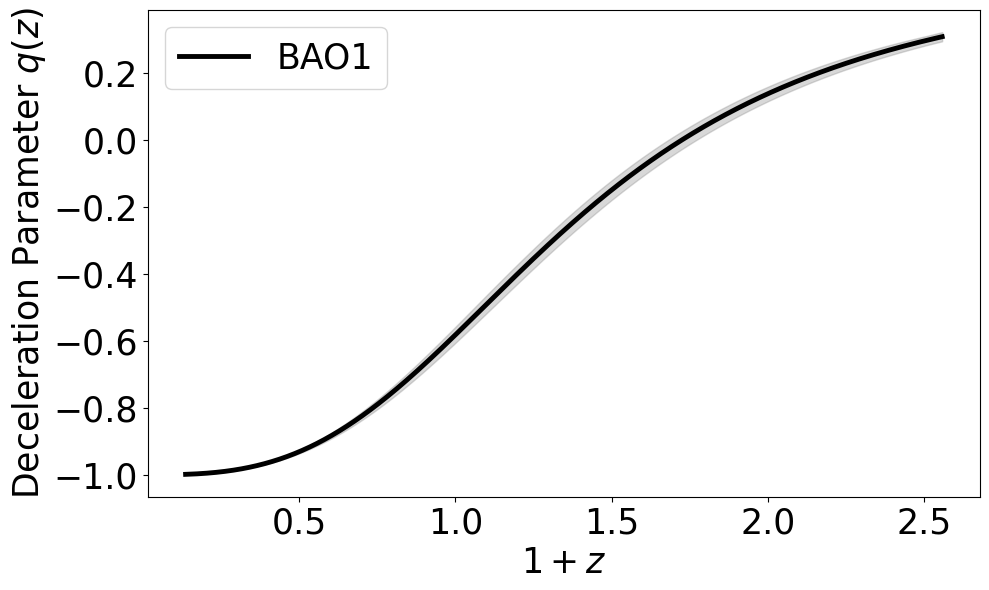}
\includegraphics[scale=0.22]{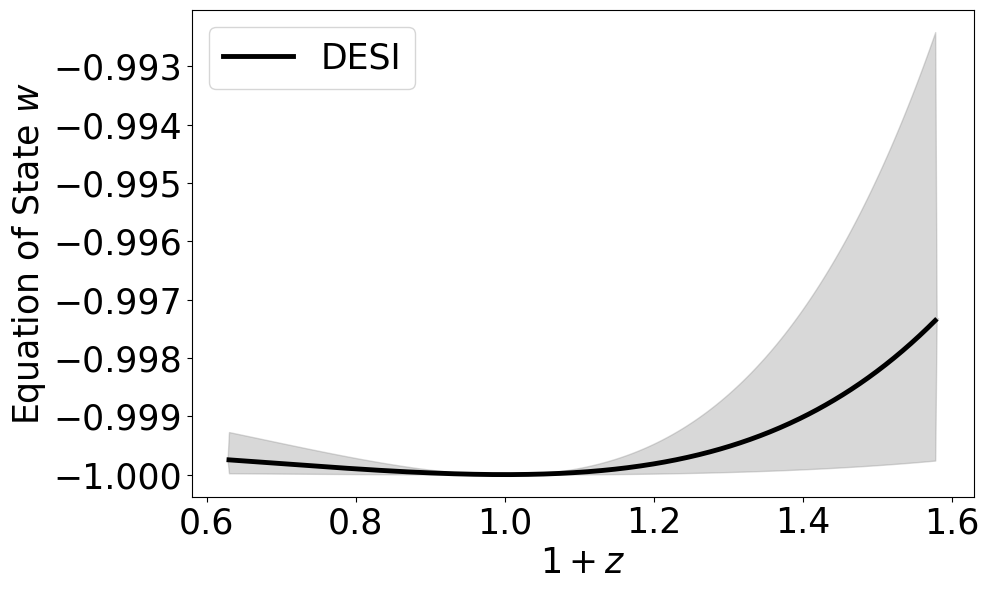}
\includegraphics[scale=0.22]{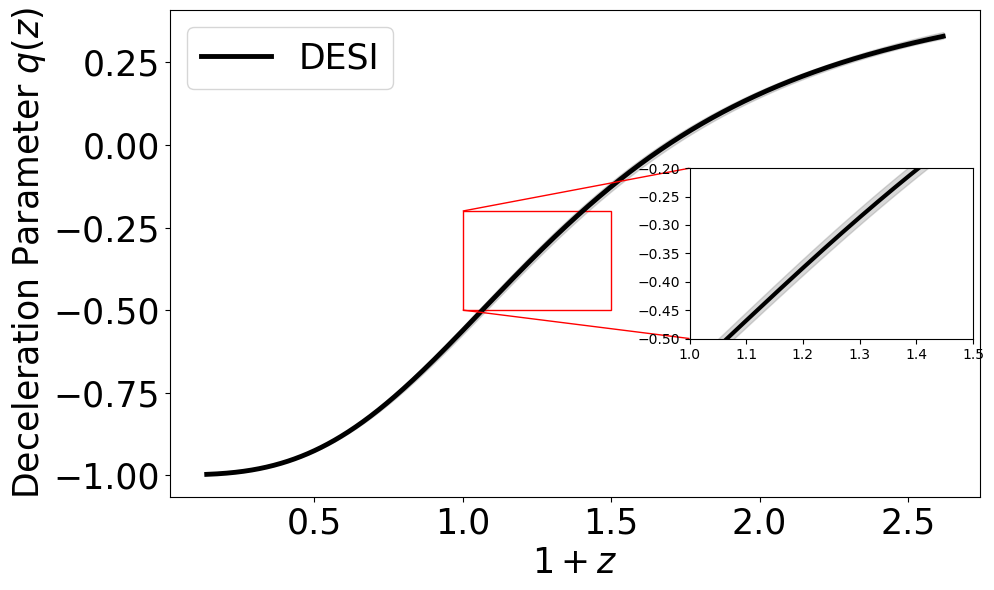}
\caption{1-$\sigma$ plots of equation of state parameter (first column) and deceleration parameter (second column). The black, solid curve represents the evolution of $w$ (first column) and $q$ (second column) as a function of redshift $z$ at the mean values of the parameters. 
The shaded region corresponds to the 1-$\sigma$ ranges of the parameters.}
\label{decelaration no w}
\end{figure}

In this model, we have a five-dimensional parameter space, comprising $H_0$, $\Omega_{m0}$, $\lambda$ and $\delta$ and $w_{\phi 0}$.     
We have imposed the uniform prior as in table \ref{Param_bound no w}.
Using this choice of priors, the parameter estimates obtained from the MCMC analysis are presented in table \ref{Param_bound w} and figure \ref{mcmc w} presents the results of the MCMC analysis.
Using the obtained parameter values, we have plotted the evolution of the equation of state parameter and the deceleration parameter in figures \ref{eq of state w} and \ref{decelaration w} respectively for all three datasets considered. The analysis will be repeated with $w_{\phi 0}$ fixed at $-1$ for reference, keeping all other priors within the same ranges (see the table \ref{Param_estimantion no w} and the figures \ref{mcmc} and \ref{decelaration no w}). {We will henceforth refer to the $w_{\phi 0} = -1$ case as the `fixed-w' model and the $w_{\phi 0} \neq -1 $ case as the `free-w model'.}
\begin{table}[t]
\centering
\begin{tabular}{|c|c|c|} 
    \hline
    Parameter &  Prior  \\ 
    \hline
    $H_0$ & $[50, 100]~km s^{-1} Mpc^{-1}$ \\ 
    \hline
     $\Omega_m $ & $[0.01, 1]$ \\ 
   \hline
    $\lambda$   &$ [0 ,1]$ \\ 
    \hline
     $\delta$   &  $[0.5,1.5]$ \\ 
     \hline
     $w_{\phi 0}$ & $[-1,0]$\\
    \hline
\end{tabular}
\caption{This table lists the uniform priors assumed for the parameters in the MCMC analysis for all datasets included in this study.}
\label{Param_bound no w}
\end{table}

\begin{table}[h]
\centering
\begin{tabular}{|c|c|c|c|c|c|c|c|c|c|c|} 
    \hline
    &  $H_0$ &$\sigma_{H_0}$ & $\Omega_{m0}$ & $\sigma_{\Omega_{m0}}$ & $\lambda$ & $\sigma_{\lambda}$ & $\delta$ & $\sigma_{\delta}$ & $w_{\phi 0}$ & $\sigma_{w_{\phi 0}}$ \\ 
    \hline
  PANTHEON+  & 70.774  & 1.659 & 0.297 & 0.065 & 0.757  & 0.211  & 0.789  & 0.266 & -0.862 & 0.089
     \\
       \hline
      BAO1 & 67.611  & 1.706 & 0.278 &  0.029 & 0.721  &  0.225 & 0.852 & 0.285 & -0.917 & 0.072
     \\
     \hline
    DESI DR2  & 66.743  &  1.845 & 0.288 & 0.034 & 0.759  & 0.208 & 0.795  & 0.270 & -0.896 & 0.083 \\
    \hline

\end{tabular}
\caption{The parameter estimates, including their mean values and standard deviations, were derived from the chains produced by the MCMC analysis.
}
\label{Param_bound w}
\end{table}

\begin{table}[ht]
\centering
\begin{tabular}{|c|c|c|c|c|c|c|c|c|} 
    \hline
    &  $H_0$ &$\sigma_{H_0}$ & $\Omega_{m0}$ & $\sigma_{\Omega_{m0}}$ & $\lambda$ & $\sigma_{\lambda}$ & $\delta$ & $\sigma_{\delta}$ \\ 
    \hline
  PANTHEON+   & 72.931  & 0.233 & 0.358 & 0.019 & 0.272  & 0.216  & 1.076  & 0.277
     \\
       \hline
      BAO1 & 69.442  & 0.748 & 0.279 & 0.016 & 0.065  &  0.046 & 1.080 & 0.278
     \\
     \hline
    DESI DR2  & 69.123  &  0.432 & 0.292 & 0.008 & 0.063  & 0.044 & 1.074  & 0.277 \\
    \hline

\end{tabular}
\caption{This table presents the best fit values and errors in the parameters derived from the MCMC chains, for the reference case in which the present-day equation-of-state parameter $w_{\phi 0}$ is fixed to -1.
}
\label{Param_estimantion no w}
\end{table}

It is important to note that 
{for free-w model}, 
all the cosmological parameters exhibit consistency within  \(\sim 1\sigma\) range across the datasets considered. This helps in reducing the disagreement in parameter estimates, particularly for the Hubble constant \(H_0\). However, this improvement comes at the cost of larger error bars in \(H_0\), especially for the DESI BAO dataset.
In contrast, 
{for the fixed-w case},
the values of \(H_0\) obtained from different datasets (table \ref{Param_estimantion no w}) continue to show significant tension. In this scenario, the DESI data provides the most stringent constraints on the deceleration parameter \(q(z)\) among all the datasets considered.

From the evolution of the equation-of-state parameter \(w_\phi\) and the deceleration parameter \(q\), it is evident that when \(w_{\phi 0}\) is included as part of the parameter space,  the dark energy component \(w_\phi\) is expected to undergo a turnaround in its equation of state not too far in the past leading to a future turnaround in the deceleration parameter  \(q\) as well. This implying that the acceleration of the universe’s expansion will eventually begin to slow down. In contrast, 
{in the fixed-w model},
the evolution inferred from the datasets suggests a future turnaround in \(w_\phi\), while the acceleration of the universe’s expansion is expected to persist, gradually approaching the asymptotic value \(q = -1\).
At redshift \(z = 0\), the central values and the corresponding \(1\sigma\) ranges of the deceleration parameter \(q\) obtained from different datasets are listed in table \ref{deceleration parameter at z=0}. These values differ from the result reported in \cite{riess2022comprehensive} by a 0.08–0.96\(\sigma\) range, representing a considerably better agreement compared to the 
{fixed-w model},
where the present-day values of \(q\) differ from the reported value \(q_0 = -0.51 \pm 0.024\) by a 1.6–3\(\sigma\) range.

\begin{table}[h!]
\centering
\begin{tabular}{|l|c|c|}
\hline
\multirow{2}{*}{\textbf{Dataset}} & \multicolumn{2}{c|}{\boldmath$q_0$} \\
\cline{2-3}
 & 
{Fixed-w model}
 & 
 {Free-w model}
 \\
\hline
& & 
\\
PANTHEON+   & $  -0.463 \pm 0.0285$ & $  -0.430^{+0.083}_{-0.081}$ \\
& & 
\\
BAO1      & $  -0.581 \pm 0.024$  & $  -0.504^{+0.069}_{-0.068}$ \\
& & 
\\
DESI DR2  & $  -0.563 \pm 0.012$  & $  -0.471^{+0.078}_{-0.077}$ \\
\hline
\end{tabular}
\caption{Central values and $1\sigma$ uncertainties of the deceleration parameter \( q(0) \) at redshift \( z = 0 \), for fixed equation of state parameter, \( w_{\phi 0} =-1\) and varying present-day \( w_{\phi 0} \) (refer values from table \ref{Param_bound w}) for the respective data.}
\label{deceleration parameter at z=0}
\end{table}

Let us now carry out an {Akaike Information criterion (AIC)} and {Bayesian Information Criterion (BIC)} analysis to compare the 
{fixed-w and free-w model},
in order to determine which model provides a better fit to the data. {Both AIC and BIC penalize complex models, but as the sample size increases, BIC penalizes more strongly than AIC, thereby favouring a  model with lesser number of parameters.}
The AIC and BIC is given by
\begin{align}
AIC & = -2log(L)+2k,\\
    BIC & = -2log(L)+k\hspace{0.5mm}log(n).
\end{align} 
{where k is the number of estimated parameters and n is the sample size.} We assume the likelihood function to be $L \propto e^{-\frac{\chi^2}{2}}$, and compare the models using the differences in their AIC and BIC values. This approach avoids the need to compute the normalization constant, which is identical for both models. 

 \begin{align}
\Delta AIC &=AIC_1-AIC_2 = \chi^2_1-\chi^2_2 +2(k_1-k_2),\label{AIC}\\
  \Delta  BIC &=BIC_1-BIC_2 = \chi^2_1-\chi^2_2+k_1\hspace{0.5mm}log(n_1)-k_2\hspace{0.5mm}log(n_2).
  \label{BIC}
\end{align} 

\begin{table}[h!]
\centering
\begin{tabular}{|l|c|c|}
\hline
\multirow{1}{*}{\textbf{Dataset}} & \multicolumn{1}{c|}{\boldmath$\Delta AIC$} &\multicolumn{1}{c|}{\boldmath$\Delta BIC$} \\
\cline{1-3}
PANTHEON+    & -1.459 &7.98\\
BAO1        &-30.091 & -31.556\\
DESI DR2     & -177.475 &-178.695\\
\hline
\end{tabular}
\caption{This table presents the differences in AIC and BIC, 
The results show that the Type Ia supernova dataset favors 
 {free-w model},
 whereas the BAO dataset favors 
 {fixed-w model}.
}
\label{AIC and BIC}
\end{table}
{In equations \ref{AIC} and \ref{BIC}, the subscript 1 is used for fixed-w model and subscript 2 is used for free-w model}. Based on the AIC and BIC analyses comparing the two models, it is found that both BAO1 and DESI BAO DR2 data favor the 
{fixed-w}
model. In contrast, the Pantheon+ supernova data shows a neutral preference according to the AIC, while the BIC indicates a preference for the 
{free-w model}.
This highlights a clear divergence in the model preferences inferred from the SNe type 1A and BAO datasets. 

\section{Conclusion \& Summary}
\label{summary}
In this work, we have presented an analysis of a tachyon-type scalar field model as a candidate for dark energy, adopting an exponential potential for the field. The Friedman equations were solved under the assumption of a spatially flat universe, neglecting the contribution from radiation. Our model involves five parameters
{and is referred to as the free-w model}, 
with the case \( w_{\phi 0} = -1 \) (the present-day value of the equation of state parameter) taken as the reference model 
{and referred to as the fixed-w model}.
The priors have been fixed as specified.
We performed Markov Chain Monte Carlo (MCMC) analyses to constrain the model parameters using Supernovae type Ia Pantheon+ data, existing BAO data (referred to as BAO1) and the recent DESI BAO DR2 data. 
Our results indicate that the parameter estimates, particularly for the Hubble constant \(H_0\), remain consistent within the \(1\sigma\) range across different datasets. However, this improvement is primarily due to the larger error bars associated with $H_0$ in the BAO datasets.
 When compared to the 
{fixed-w model},
 the discrepancy in parameter values, especially in \(H_0\), continues to persist across the datasets. Among them, the DESI dataset yields the most stringent constraints in this case.

Using the best-fit parameters and their associated $1~\sigma$ error bars, we analysed the evolution of both the equation of state parameter $w_{\phi}$ and the deceleration parameter $q$, and compared them with the reference model.
From the MCMC analysis, we found that the present-day value of the equation of state parameter lies between $-0.86$ and $-0.92$ across the datasets considered. The evolution of the equation of state indicates that a turnaround could occur not too far in the past, provided the other parameters are suitably fine-tuned. A similar trend is observed for the deceleration parameter, indicating that the universe may currently be transitioning into a phase where the acceleration of its expansion is gradually slowing down. The present-day value of the deceleration parameter in this case agrees with the SH0ES data within the $0.08\sigma$–$0.96\sigma$ range.
In contrast, for the 
{fixed-w model},
our analysis indicates that the possible turnaround in the equation of state parameter would occur in the future. In this case, the present-day value of the deceleration parameter agrees with the value reported by the SH0ES data within the $1.6\sigma$–$3\sigma$ range and is expected to asymptotically approach the value of $-1$ in the future. Thus, the expansion cannot slow down; rather, the acceleration keeps increasing and $q$ eventually approaches \(-1\), where it is expected to settle in the future.
However, when the model preference is examined using AIC and BIC analyses, we find that the Type Ia supernova dataset favours 
{free-w model},
whereas the BAO datasets prefer the 
{fixed-w model}.
This means that, within this framework, both datasets agree that the equation of state will undergo a future turnaround. However, they diverge in their predictions for the deceleration parameter. The SNe type Ia dataset select parameters that lead to the slowing of the universe’s expansion in the future epoch, while the BAO data chooses parameters that produce increasing acceleration, with $q$ asymptotically approaching \(-1\). This highlights a clear disagreement in the future cosmological evolution predicted by the two datasets within the tachyon-type dark energy model.

\FloatBarrier
\section*{Acknowledgment}
The authors thank J. S. Bagla for useful discussion. Athul C N thanks IIT Kanpur for hospitality, as this manuscript was completed during his term as Project Associate-1. We acknowledge IISER Mohali for providing computing resources of High Performance Computing Facility at IISER Mohali.

\appendix

\medskip
\bibliographystyle{JHEP}
\bibliography{EPJC}

\end{document}